\documentclass[12pt,epsfig,citesort,svgnames,dvipsnames]{article}
\usepackage{multicol}
\usepackage[all]{xy}
\usepackage{graphicx}
\usepackage{todonotes}
\usepackage{amsmath}
\usepackage{amssymb}
\usepackage{amsfonts}
\usepackage{amsthm}

\theoremstyle{definition}

\theoremstyle{remark}

\usepackage{amscd}
\usepackage{latexsym}
\usepackage[official]{eurosym}
\usepackage[english]{babel}
\usepackage[T1]{fontenc}
\usepackage[utf8]{inputenc}
\usepackage{authblk}
\usepackage{natbib}
\usepackage{xcolor}
\usepackage{hyperref}
\usepackage[nottoc]{tocbibind}
\hypersetup{
colorlinks=true,
citecolor=WildStrawberry,
anchorcolor=Black,
filecolor=cyan,
menucolor=magenta,
runcolor=cyan,
linkcolor=RoyalBlue,
linktoc=page, 
urlcolor=TealBlue}

\title{\bf {A Proposal for a Metaphysics of Self-Subsisting Structures.\\ II. Quantum Physics}}
\author[a]{\normalsize Antonio Vassallo
\thanks{antonio.vassallo1977@gmail.com, ORCID 0000-0002-1846-5362}}
\author[b]{\normalsize Pedro Naranjo}
\author[c]{\normalsize Tim Koslowski}
\affil[a]{\normalsize \emph{Warsaw University of Technology, Faculty of Administration and Social Sciences, Plac Politechniki 1, 00-661 Warsaw, Poland}}
\affil[b]{\normalsize \emph{Plaza Mayor, 4/1B, 09003 Burgos, Spain}}
\affil[c]{\normalsize  \emph{Technical University of Applied Sciences W\"urzburg-Schweinfurt, Faculty of Applied Natural Sciences and Humanities, M\"unzstr. 12, 97070 W\"urzburg, Germany}}

\date{}

\begin{document}
\maketitle
%\vspace{-5mm}
\begin{center}
Forthcoming in \emph{Foundations of Physics}.
\end{center}
\begin{abstract}
The paper presents an extension of the metaphysics of self-subsisting structures set out in a companion paper to the realm of non-relativistic quantum physics. The discussion is centered around a Pure Shape Dynamics model representing a relational implementation of a de Broglie-Bohm $N$-body system. An interpretation of this model in terms of self-subsisting structures is proposed and assessed against the background of the debate on the metaphysics of quantum physics, with a particular emphasis on the nature of the wave function. The analysis shows that elaborating an appropriate Leibnizian/Machian metaphysics of the quantum world requires a substantial revision of the notion of world-building relation.
\end{abstract}
\newpage
\tableofcontents

\section{Outline}
The paper presents an extension of the metaphysics of self-subsisting structures for Pure Shape Dynamics (PSD) to the realm of non-relativistic quantum physics. As such, it relies on the groundwork laid down in the companion paper \citet{vasnarkos2}, as well as the technical results reported in \citet{726} and \citet{747}. To keep the discussion as much self-contained as possible, a brief presentation of the overall framework will be provided in section \ref{sec:int}. Section \ref{sec:dbb} will present a simple de Broglie-Bohm model of PSD, highlighting its pivotal features and the subtleties involved in implementing it. Section \ref{subsec:nopsi} will discuss two possible metaphysical choices compatible with the model, arguing that a metaphysics of self-subsisting structures is the better way to go. Section \ref{sec:wf} will consider the role of the wave function in determining the quantum behavior of self-subsisting structures and what its place in the ontology is (if any). Section \ref{subsec:PO} will discuss whether this PSD model represents an implementation of the so-called \emph{primitive ontology} approach to quantum physics. Finally, section \ref{sec:disc} will put the preliminary results presented in this paper in perspective via some concluding remarks.

\section{Introduction: Pure Shape Dynamics as a Theory of Self-Subsisting Structures}\label{sec:int}

\subsection{Leibnizian/Machian Relationalism}

The modern relational approach to physics---pioneered by Julian Barbour and Bruno Bertotti \citep{83}---seeks to reach a complete description of a closed physical system in purely intrinsic terms, without reference to anything external to it. While this attitude, intended as a pragmatic choice, sounds uncontroversial to the point of triviality---for example, when describing isolated molecules \citep[see, e.g., the review in][]{litren}---, it assumes heavy metaphysical undertones when interpreted in a Leibnizian/Machian spirit. In this latter case, the elimination of external reference structures should be intended as a means to eschew from the ontology of the physical world any degree of freedom that is not inherent in material happenings, spatiotemporal degrees of freedom \emph{in primis}. This inclination to ontological parsimony (and despise of external spatiotemporal structures) is rooted in a comparativist take on the definition of physical quantities: Since every measurement consists in comparing a physical magnitude with a chosen standard unit, only ratios of physical quantities carry objective information (i.e., are physically observable). Therefore, it makes no sense to postulate unobservable physical magnitudes whose variation is not defined by way of comparison to something else---contrast this with Newtonian time, which flows independently of anything else in the universe. This is, in a nutshell, the spirit of Leibnizian/Machian relationalism: The objective and complete physical description of any part of a closed system has to be given in reference to each and every other part of it. This is particularly true of the spatiotemporal notions associated with such a description:

\begin{quote}
[L]eibniz and Mach suggest that if we want to get a true idea of what a point of space-time is like we should look outward at the universe, not inward into some supposed amorphous treacle called the space-time manifold. The complete notion of a point of space-time in fact consists of the appearance of the entire universe as seen from that point. Copernicus did not convince people that the earth was moving by getting them to examine the earth but rather the heavens. Similarly, the reality of different points of space-time rests ultimately on the existence of different (coherently related) viewpoints of the universe as a whole.\\
(\citealp{77}, p.265)
\end{quote}

Note how this Leibnizian/Machian approach introduces a universal perspective on physics: Full knowledge about local matters of fact is attained only when considering how such facts are related to the rest of the universe. The paradigmatic example of this attitude is the Machian perspective on inertia. Most famously, Mach replied to Newton's bucket argument by pointing out that the inertial effects felt by the water contained in a spinning bucket do not have a local origin---i.e., they are not due to the water/bucket relative motions---but should be referred to the rotation with respect to the distant ``fixed stars'' (see, e.g., \citealp{483b}, \S1, for a nuanced discussion of Mach's ideas on inertia).

The technical implementation of modern Leibnizian/Machian relationalism has steadily evolved since Barbour and Bertotti's original proposal (see, e.g., \citealp{vasnarkos1}, \S4.2, for a quick overview of this process). The relational framework endorsed in this paper is that of PSD, a natural evolution of \emph{Shape Dynamics} (SD; see \citealp{726}, for a technical introduction and \citealp{vasnarkos2}, \S2.2, for a comparison with SD). In a nutshell, PSD implements a totally intrinsic description of a physical system in two steps. First, all the redundancies associated with an external space are removed via a procedure that quotients out all the appropriate unobservable degrees of freedom---in the classical case, for example, this amounts to quotienting out the degrees of freedom associated with rigid translations, rotations, and dilations. This procedure translates the non-relational configuration space into a new one, called \emph{shape space}, $\mathcal S$. Second, the elimination of unobservable temporal degrees of freedom is obtained by demanding that the history of the physical system in shape space be described in terms of an \emph{unparametrized} curve $\gamma_0$. Intuitively, this means that the different steps in the evolution are individuated by the intrinsic degrees of freedom of the system, not by a label arbitrarily placed on $\gamma_0$ from the outside.

The major novelty of the relationalist picture advocated by PSD is exactly its insistence on using only \emph{intrinsic} geometric properties of $\gamma_0$ in $\mathcal S$ in the description of the evolution of a given physical system, which is expressed in terms of the \emph{equation of state} of $\gamma_0$:

\begin{equation}
\begin{array}{rcl}
   dq^a&=&u^a(q^a,\alpha _I^a), \\
   d\alpha _I^a &=&\Omega _I^a(q^a,\alpha_I^a)\,,
   \end{array}
   \label{curve0}
\end{equation}

where we demand that the right-hand side be described in terms of dimensionless and scale-invariant quantities, whose intrinsic change is obtained employing Hamilton's equations of motion. In \eqref{curve0}, $q^a$ are points in shape space, namely they represent the universal configurations of the system, $u^a$ is the unit tangent vector defined by the shape momenta $p_a$: 
\begin{equation}
    u^a\equiv g^{ab}(q)\frac{p_b}{\sqrt{g^{cd}p_cp_d}}\,,
    \label{unittangent}
\end{equation}

which allows us to define the direction $\phi ^A$ at $q^a$ ($g^{ab}$ being an appropriate metric defined on shape space, which will be discussed in a moment). It is through the unit tangent vector and the associated direction that the shape momenta enter Hamilton's equations, which are in turn used in the intermediary steps leading to the equation of state \eqref{curve0}. Finally, $\alpha _I^a$ is the set of any further degrees of freedom needed to fully describe the system. It is this set $\alpha _I^a$ that includes higher-order derivatives of the curve and spares us of the need of additional non-shape degrees of freedom, unlike standard SD (see, again, \citealp{vasnarkos2}, \S2.2). Among these, one parameter definitely stands out: A measure of the deviation of the curve from geodesic dynamics (cf. $\kappa$ in \S\ref{sec:dbb}, equation \eqref{ukappadefinition}).
For consistency, the elements in $\alpha _I^a$ in \eqref{curve0} must exhaust the set of all possible dimensionless and scale-invariant quantities that can be formed out of the different parameters entering a given theory. Remarkably enough, as shown in \citet{726}, the equation of state should be taken as a whole and interpreted as giving the relative rates of change of the degrees of freedom of the curve in shape space. Thus, one has $\frac{dq^a/ds}{d\alpha _I^a/ds}=\frac{dq^a}{d\alpha _I^a}$ (the arc-length parameter $s$ dropping out), effectively rendering the dynamics explicitly unparametrized, hence justifying the form adopted by \eqref{curve0}. 

\subsection{Two Simple Classical Models of Pure Shape Dynamics}

For the sake of illustration, we will consider two models and their associated equations of state of the respective unparametrized curves in shape space. The first one describes geodesic dynamics in shape space. This is the simplest, albeit admittedly uninteresting, physical model one can consider. The second model, the relational $N$-body system, will explicitly exhibit deviation from geodesic dynamics, thereby rendering the ensuing dynamics able to account for structure formation. Furthermore, this discussion will clearly show one of the central tenets of the program, namely, its use of only intrinsic geometric properties $\alpha _I^a$ of the curve. In a nutshell, the size of the set $\alpha _I^a$ depends on the complexity of the system to be described: The more complex the system, the more geometric properties are required in its physical description. 

\subsubsection{Geodesic System}

The dynamical system described by equation \eqref{curve0}, with the tangential direction $\phi ^A$ at the point $q^a$ as the only element in the set $\alpha _I^a$, corresponds to the simplest system in shape space, namely one whose curve is a geodesic. It is given by the Hamiltonian:

\begin{equation}
        H=\frac{1}{2}g^{ab}(q)p_a p_b\,,
        \label{Hamiltoniangeod}
    \end{equation}
where $g^{ab}$ is an appropriately defined metric on shape space, called \emph{kinematic} because---as equation \eqref{Hamiltoniangeod} suggests---it is part of the definition of the kinetic energy of a physical system. Absorbing the dimensionless mass ratios $\mu_i:=\frac{m_i}{M}$ into the configuration space metric (see \eqref{kmetric} below) allows us to set the overall mean mass $M:=\frac{1}{N}\sum_{i=1}^N m_i$ to unity. In order to work out the equation of state, let us conveniently use the arc-length parametrization of the curve with respect to $g_{ab}(q)$:

\begin{equation}
 \left(\frac{ds}{dt}\right)^2=g_{ab}(q)\frac{dq^a}{dt}\frac{dq^b}{dt}=g^{ab}(q)p_ap_b\,,
 \label{arc-length}
\end{equation}
where $s$ is the arc-length parameter, which readily yields

\begin{equation}\label{equ:KinematicEOM}
 \frac{dq^a}{ds}=g^{ab}(q)\frac{p_b}{\sqrt{g^{cd}(q)p_cp_d}}\,,
\end{equation}
where the right-hand side is the unit tangent vector $u^a$, \eqref{unittangent}, which, recall, defines the direction $\phi ^A$ at $q^a$. Taking any explicit functional expression $\Phi (q,p)$ for the direction, one gets: 

\begin{equation*}
\frac{d\phi ^A}{ds} = \frac{\partial \Phi}{\partial q^a}\,\frac{dq^a}{ds} + \frac{\partial \Phi}{\partial p_a}\,\frac{dp_a}{ds}\,,
\end{equation*}
which, by means of \eqref{unittangent}, \eqref{arc-length} and \eqref{equ:KinematicEOM}, enables us to write the equation of state of the geodesic curve by means of the canonical equations of motion generated by the Hamiltonian \eqref{Hamiltoniangeod} as:

\begin{equation}
 \begin{array}{rcl}
   d\,q^a &=& u^a(q,\phi)\,,\\
   d\,\phi ^A&=& \frac{\partial \Phi}{\partial q^a}\,u^a(q,\phi)-\frac 1 2\frac{\partial \Phi}{\partial u^a} g^{bc}_{\phantom{bc},a}(q)u_b(q,\phi)u_{c}(q,\phi)\,,
 \end{array}
 \label{geodesic}
\end{equation}
where we have dropped the arc-length parameter $s$ to emphasize the underlying unparametrized nature of the curve.

In general, geodesic motions over shape space are physically uninteresting because they do not lead to the formation of any stable subsystems. An example is a perpetually expanding classical $N$-body system, where the inter-particle separations grow indefinitely. %The next model will clarify why the formation of stable subsystems is of paramount importance to describe realistic physical models. 

%It is worth stressing one crucial point about geodesic dynamics on \emph{compact} configuration spaces, which is the case of the associated shape space here: Poincaré's recurrence theorem implies that the dynamics will not be able to yield formation of stable records, which is of paramount importance to the analysis of the arrow of time, as will be stressed in the next model.

\subsubsection{Newtonian $E=0\,\,\,N$-body System}\label{sec:ephs}
% (see \citealp{726})
Let us consider the Hamiltonian with a generic potential $V$:
    \begin{equation}
        H=\frac{1}{2L^2}(D^2+g^{ab}(q)p_a p_b)+V(L,q)\,,
        \label{HamiltonianNbody}
    \end{equation}
where the overall mean mass $M:=\frac{1}{N}\sum_{i=1}^N m_i$ has been absorbed into the coupling constant of the potential, \{$q^a,p_a$\} are coordinates and momenta in shape space, and $D$ is the dilatational momentum.\footnote{More precisely, $D=\sum _1^N \mathbf{r}_a^{\mathrm{cm}}\,\cdot \mathbf{p}^a_{\mathrm{cm}}$, where $\mathbf{r}_a^{\mathrm{cm}}$ and $\mathbf{p}^a_{\mathrm{cm}}$ are, respectively, the position and momentum of particle $a$ relative to the system's center of mass.} Moreover, $g^{ab}$ is the kinematic metric, which, in the case of Euclidean space, is given by the scale-free Euclidean metric on configuration space:
\begin{equation}\label{scalefac}
 ds^2=\frac{\sum_{I=1}^N\,d\bf r_I^2}{L^2}\,,
\end{equation}
where $L^2:=\sum_{I=1}^N\bf r_I^2$ denotes the square of the total scale in the center-of-mass frame. In ``global'' coordinates, the induced kinematic metric on shape space takes the explicit form: 
\begin{equation}\label{kmetric}
 \frac{1}{L^2}\,g^{ab}(q):=
 %\sum_{I,J=1}^N\sum_{m,n=1}^3
 g_{IJ}\frac{\partial q^a}{\partial r^I}\frac{\partial q^b}{\partial r^J}\,,
\end{equation} 
 where $r^I$ and $g_{IJ}$ ---which contains the dimensionless mass ratios $\mu_i:=\frac{m_i}{M}$ --- are some coordinates and metric, respectively, in configuration space. We shall focus on the case of a homogeneous potential, $V(L,q)=\beta L^k C(q)$, with $\beta$ an arbitrary coupling constant into which the overall mass $M$ above has been absorbed, which enables us to write \eqref{HamiltonianNbody} as an energy conservation constraint: 
 \begin{equation}
        H=\frac{1}{2}(D^2+p^2)+\beta L^{k+2}C(q)=0\,,
        \label{HamiltonianNbodyhomog}
    \end{equation}
where $p\equiv \sqrt{g^{ab}p_a p_b}$ is the length of the shape momenta. Unlike the geodesic system \eqref{geodesic}, now one can build a further degree of freedom in $\alpha _I^a$, namely $\kappa\equiv\frac{p^2}{\beta L^{k+2}}$, which is related to $K^{-1}$. $K$ is the curvature of the curve, which, is given in terms of the acceleration vectors (see \citealp[\S 3.4]{726} for details). Thus, $\kappa$ can be thought of as an \emph{intrinsic acceleration}, i.e., a measure of how much the curve traced out by a given physical system deviates from geodesic dynamics, \eqref{geodesic}. The intrinsic change of $\kappa$ is obtained, once again, by Hamilton equations of motion, yielding the following equation of state: 
\begin{equation}
 \begin{array}{rcl}
  dq^a&=&g^{ab}(q)u_b(q,\phi)\\
  d\phi ^A&=&\frac{\partial\Phi_A}{\partial q^a}\,g^{ab}(q)u_b(q,\phi)-\frac{\partial \Phi_A}{\partial u_a}\left(\frac{1}{\kappa}\frac{\partial C(q)}{\partial q ^a}+\frac 1 2 {g^{bc}}_{,a}(q)\,u_b(q,\phi)u_c(q,\phi)\right)\\
  d\kappa&=&-(k+2)\kappa\varepsilon(q,\phi,\kappa)-2u^a(q,\phi)C_{,a}(q)\,,
 \end{array}
 \label{NbodyzeroE}
\end{equation}
where $\varepsilon\equiv\frac{D}{p}=\pm\sqrt{-\left(1+2\frac{C(q)}{\kappa}\right)}$ can be solved for in terms of \{$q^a,\phi ^A,\kappa$\} by means of the energy conservation constraint \eqref{HamiltonianNbodyhomog} and $C(q)$ is the \emph{shape potential} which, roughly, measures the amount of clustering in the system and, hence, accounts for the formation of (stable) subsystems. %function (see \citealp{706} for the original paper introducing this function and \citealp[][\S 3.5]{726} for the discussion within the PSD framework). 

%Remarkably enough, the dynamical system \eqref{NbodyzeroE}, with the key role played by $\kappa$, provides a compelling case for the emergence of the arrow of time: by sidestepping Poincaré's recurrence theorem, this system yields stable structure formation as a consequence of the attractor-driven behaviour of complexity in shape space.

\subsection{Emergence of Scale and Duration}

How are the standard Newtonian notions of scale and duration be recovered from \eqref{NbodyzeroE}? Although it is clear that it will be impossible to get absolute units of scale and duration, since the curve $\gamma_0$ lacks any such structures, one can nonetheless meaningfully ask how definitions of absolute scales evolve. Let us consider a curve $\gamma$ in shape space and two points $q^a$ and $q^b$ on it. Next, we will define the total scale $L$ of the system at point $q^b$ to be the unit of size $L_0$ and the total duration between $q^a$ and $q^b$ to be the unit of time $T$. Then, given a third point $q^c$ on $\gamma$, one may ask: What is the total scale $L$ measured in units of $L_0$? Likewise, what is the duration between $q^b$ and $q^c$ in units of $T$? To answer these questions, we will give explicit equations for standard scale and duration in the case of the homogeneous system \eqref{HamiltonianNbodyhomog}. Following the suggestion in \citet[][\S 4]{135}, we will call these standards \emph{ephemeris scale} and \emph{ephemeris duration}. Historically, the ephemeris time was a duration standard used by astronomers and based on the intrinsic properties of the solar system (considered as a closed system). More precisely, the ephemeris time was the duration standard which made all the motions of the dynamically relevant bodies compatible with Newtonian dynamics (see also \citealp[][\S 13.2.4]{514}, for a discussion of ephemeris scale and time in the context of SD).

Using the arc-length parametrization condition \eqref{arc-length}, we obtain the ephemeris scale equation (cf. \citealp[][\S 3.6]{726}, for the technical details of the derivation): 
\begin{equation}
  \frac{d}{d\,s}\log L=\frac{D}{p}=\pm\sqrt{-\left(1+2\frac{C(q)}{\kappa}\right)}\,.
  \label{ephemerisScale}
\end{equation}

Notice that the shape potential $C(q)$ in Newtonian gravity is negative definite, which must be taken into account in \eqref{ephemerisScale}.

Likewise, the ephemeris duration equation is derived from
\begin{equation}
  \frac{d}{d\,s}\log\left(\frac{ds}{dt}\right)=\frac{d}{d\,s}\log p= -\frac{2}{\kappa}u^a(q,\phi)\,C_{,a}(q)\,.
  \label{ephemerisDuration}
\end{equation} 
The unit of time $T$ is obtained by integrating \eqref{ephemerisDuration} between configurations $q^a$ and $q^b$. 

As expected, the right-hand sides of both \eqref{ephemerisScale} and \eqref{ephemerisDuration} refer only to intrinsic properties of the unparametrized curve in shape space. In particular, the (rate of change of) $C(q)$ gives rise to the standard, global notion of duration, exhibiting the general result of the emergence of an arrow of time in terms of the increase in complexity of the $N$-body system (see \citealp{706} for the original paper introducing this result and \citealp[][\S 3.5]{726}, for the discussion within the PSD framework). 

It is important to point out that (i) the ephemeris equations are model-dependent and (ii) there exists a many-to-one correspondence between Newtonian models and equations of state of curves in shape space, so the ephemeris equations are in general not uniquely associated with an equation of state of the curve in shape space. 

\subsection{General Remarks}
After these two examples to illustrate the general PSD framework, some important comments are in order. Firstly, it is clear from the above discussion that the fundamental structure of $\mathcal S$ describing the equation of state of $\gamma_0$---and, hence, the evolution of a physical system---is largely topological. This topological nature in turn guarantees that the dynamics does not need any parametrization---even if some parametrization $s$ may still be used to simplify the computations. Secondly, notwithstanding the previous point, some minimal geometrical structure is certainly needed to describe the curve, namely a metric $g_{ab}$ on shape space. Two facts are worth stressing: (i) A metric on shape space does \emph{not} measure size, but the amount of similarity in the configurations, and (ii) the naturally induced metric on shape space is the already mentioned kinematic metric. Thirdly, note also that dynamical parameters, such as mass, charge, and spin, enter the equation of state through \emph{scale-invariant} and \emph{dimensionless} quantities. Hence, the conformal structure alone does not individuate any dimensionful quantity. Thus, it may be argued that dimensionful dynamical parameters and physical units are not needed by the theory and serve only as extra descriptive information about the physical system.

\subsection{The Metaphysics of Pure Shape Dynamics}

In order to elaborate a metaphysical framework that fits PSD, some key aspects of the above sketched formalism have to be noted. First of all, the quotienting out procedure can be intended as a ``parsimonization'' of the starting non-relational ontology, whereby all fundamental physical aspects that are not definable only in terms of ratios of physical magnitudes---including intrinsic properties and primitive identity---are left out of the metaphysical picture. As such, the quotienting out procedure can be said to determine the \emph{structural} properties of physical configurations (i.e., shapes) in the sense that only a ``web'' of physical relations among suitably defined \emph{relata} remains at the end of the procedure. In the case of a Newtonian $N$-particle system, for example, the starting fundamental ontology consisting of individual particles inhabiting a Euclidean 3-space is ``parsimonized'' into a structure of Euclidean spatial (conformal) relations: Being a ``material particle'' in this latter case means just being a \emph{relatum} in this structure. 

Secondly, the equations \eqref{curve0} are intended as a law schema that encompasses all possible domains of physics (classical, relativistic, and quantum). The more complex the motions to be described, the larger the amount of geometric degrees of freedom included in $\alpha _I^a$. This fact highlights a structural continuity across the model spectrum of PSD: All shapes, independently of their nature and complexity, share some basic structural tracts abstracted via the mathematical relations encoded in \eqref{curve0}. Moreover, it is obvious that PSD by construction remains invariably a theory of shapes, whatever the physical domain considered.

In short, PSD encodes the core tenets of scientific structuralism: Physics is not about collections of individual objects but about the relational, structural aspects of the world (cf. \citealp{675}, p.~15, for one of the most famous defenses of this view). The empirically adequate structural features of a physical theory are then carried over to the superseding theories, thus justifying the persistence of certain concepts throughout the theoretical advancements of physics. However, it is clear from the above discussion that PSD's structural spirit is not just epistemic but \emph{ontic}: PSD is a theory about structures; more precisely, \emph{self-subsisting} structures, given that they do not need an external embedding space in order to be individuated and characterized.

We refer the interested reader to the companion paper \citet{vasnarkos2} for a thorough presentation and discussion of the self-subsisting structures framework in a classical setting. In the remainder of the present paper, we will tackle the tricky question of how this framework looks like in the context of quantum physics and what exactly it has to contribute to some outstanding debates regarding the metaphysics of the quantum world. 

\section{de Broglie-Bohm $N$-body System}\label{sec:dbb}

\subsection{Motivation}
As already hinted at when discussing the law schema \eqref{curve0}, PSD is in principle able to account for non-relativistic quantum systems in quite the same vein as it does for classical ones. This becomes evident when considering de Broglie-Bohm systems, which roughly consist of an $N$-particle system moving under the ``influence'' of a pilot wave. The motivation for focusing on the de Broglie-Bohm theory rather than on standard quantum mechanics boils down to the fact that the former theory explains the appearance of definite measurement outcomes by postulating a concrete configuration of material particles occupying definite spatial positions at all times. In this sense, the de Broglie-Bohm theory accords a privileged ontological status to spatial concepts, which is one of the central principles of PSD. Recall that the measurement process amounts to the comparison of two subsystems, and this process involves only angles as objective empirical information. Moreover, the de Broglie-Bohm theory is akin to a Leibnizian/Machian spirit in that it introduces a ``global'' perspective on the dynamics. Indeed, this theory is first and foremost about \emph{universal} configurations of particles. From this all-encompassing description, the behavior of subsystems can be recovered via the definition of appropriate \emph{effective} wave functions whose evolution obeys a Schr\"odinger-like evolution (the technical details can be found, e.g., in \citealp[][\S5]{222}). This is similar to the way classical Barbour-Bertotti theories recover Newtonian dynamical equations---including their symmetries---from the global relational description when the focus is shifted to local subsystems \citep[][\S3]{83}.

Since the de Broglie-Bohm theory is empirically equivalent to standard non-relativistic quantum mechanics (see, again, \citealp{222}, for a justification of this claim), PSD is able to recover all the empirical predictions of this latter theory (see the discussion after \eqref{debohm} for a critical assessment). The key point that makes this possible is the fact that the dynamics of the de Broglie-Bohm theory, leaving aside its inherent non-locality, exhibits the same spatiotemporal symmetries of classical mechanics. That is, both theories are Galilean invariant (but see \citealp{409}, for a dissenting voice). Since the de Broglie-Bohm dynamics has a well-defined symmetry group (modulo some caveats spelled out, e.g., in \citealp{468}), its dynamics is amenable to the quotienting out procedure mentioned in section \ref{sec:int}.

\subsection{Technical Implementation}

Given that the de Broglie-Bohm theory describes the motion of a universal configuration of particles whose collective behavior is ``choreographed'' by the wave function associated with the system, a full specification of the dynamics requires two equations: (i) The guidance equation accounting for the behavior of configurations in terms of the wave function and (ii) Schr\"odinger's equation describing the dynamics of the wave function itself. So, the quantum challenge for PSD can be recast in the following terms. PSD's focus on the geometric properties of a curve in shape space can very well be thought of as an intrinsic rendering of the guidance equation: We may, in principle, describe the change in configurations employing a (possibly infinite) set of degrees of freedom. However, the guidance equation depends on an element---the wave function---that is itself subject to dynamics, which means that also this latter element should be subjected to the quotienting out procedure that strips all ``absolute'' degrees of freedom associated with a background space out of the description. Fortunately, as we shall see in a moment, the shape space of a de Broglie-Bohm $N$-body system permits to define this ``stripped down'' counterpart of the standard pilot wave.

In order to comply with the PSD tenets, namely, a dimensionless and reparametrization invariant description of the dynamics of the de Broglie-Bohm $N$-body system, let us first briefly recall how the relevant shape space is obtained. In the standard, non-relational theory, our $N$-body system lives in an external $3$-dimensional Euclidean space $\mathbb{R}^3$, so we must identify all the configurations related by transformations that belong to the so-called {\emph{similarity group}} $\mathsf {Sim}$, namely the joint group of rigid translations $\mathsf T$, rotations $\mathsf R$, and dilations $\mathsf S$. Thus, the $N$-body shape space is $\mathcal S :=Q^N/\mathsf T\mathsf R\mathsf S$, with $Q^N$ the standard configuration space. This completely mimics the construction of the classical case. However, as we just said, the quantum counterpart comes with an additional ingredient: The wave function $\Psi$. In order to work with a consistent dynamical system, the wave function must be defined on $\mathcal S$. Physically, this simply means that all information encoded in $\Psi$ regarding the embedding of the $N$-body system in an external space must be dropped. Mathematically, this is readily achieved as follows: If $\Psi _N$ is the wave function living in $Q^N$, its restriction to shape space is $\Psi _{\mathcal S}\equiv\,\Psi _N|_{\mathsf {Sim}}$ (strictly speaking, this reduction works only if the original wave function $\Psi _N$ is single-valued along orbits of $\mathsf {Sim}$; we shall gloss over this, given it will not affect our analysis). From now on, we will omit the label $\mathcal S$ from $\Psi$ and work with wave functions on shape space. 

Equipped with this, consider the equation of state of the curve representing the succession of instantaneous configurations of the system, which are pairs $(Q^a,\Psi(q))$ of shapes and wave functions on shape space, with $Q^a$ standing for the actual configuration of the $N$-body system and $q^a$ for an arbitrary coordinate in shape space. As shown in \citet{747}, once the wave function is decomposed into its amplitude and phase (i.e., $\Psi (q) =R(q)\,e^{iS(q)}$), the equation of state of the unparametrized curve in shape space reads:\footnote{As already emphasised in the discussion following \eqref{curve0}, the equation of state has a global character, namely the left-hand side should be read as $\frac{dq^a/ds}{d\phi^A/ds}=\frac{dq^a}{d\phi ^A}$, and so on (the arc-length parameter $s$ drops out), hence the absence of the mathematical differential symbol in the right-hand sides.}

\begin{equation}\label{debohm}
 \begin{array}{rcl}
  d Q^a &=& u^a(\phi)\\
  d \phi^A &=& \frac{\partial \Phi_A}{\partial Q^a}u^a(\phi)-\frac{\partial \Phi_A}{\partial u^a} \left(\frac 1 2 g^{cd}_{,a}(Q)u_c(\phi)u_d(\phi)+\frac 1 \kappa V_{T,a}(Q)\right)\\
  d \kappa &=& -2u^a(\phi)V_{T,a}(Q) + \tilde{K}(\kappa,\alpha,Q,\phi)\\
  d R(q) &=& -\frac{1}{\sqrt{\kappa}}\left( g^{ab}(q)R_{,a}(q)S_{,b}(q)+ \frac{1}{2} R(q)\Delta S(q)\right)\\
  d S(q) &=& -\frac{1}{\sqrt{\kappa}}\left(\frac{1}{2}g^{ab}(q)S_{,a}(q)S_{,b}(q)+V_T(q)\right),
 \end{array}
\end{equation}
where $V_T(q)=C(q)-k\frac{\Delta\,R(q)}{2\,R(q)}$ is the total potential, with $C(q)$ being the classical component (the shape potential in \ref{sec:ephs}) and $V_q\equiv-\frac{\Delta\,R(q)}{2\,R(q)}$ its scale-invariant quantum component. As discussed in \citet[\S 3.1]{747}, given the absence of Planck's constant in our model, a dimensionless coupling $k$ must be introduced to enable us to have a grasp of the relative strengths of the two components. The physical role of $k$ is to determine the Bohr radius of the ground state of approximately isolated two-body subsystems. 

Note that we have used the kinematic metric $g_{ab}(q)$ on shape space to split $S^1_a :=\left.\nabla_a\,S(q)\right|_{q^a=Q^a}$ into directions $\phi^A$ ---with any local coordinates $\phi ^A$ on the fibre of the unit tangent bundle defining a set of phase space functions $\Phi_A(q,u)$--- and the additional degree of freedom $\kappa$. As already stressed above, this latter degree of freedom plays a key role in PSD, namely, it is a measure of the deviation of the dynamics from geodesic motion (recall both our general description following \eqref{curve0} and the explicit case of the $N$-body system). We thus have:

\begin{equation}\label{ukappadefinition}
 \begin{array}{rcl}
  u^a(\phi)&=&\frac{S^1_a}{\sqrt{g^{ab}(Q)S^1_aS^1_b}}\,,\\
  \kappa &=&\frac{g^{ab}(Q)S^1_aS^1_b}{L^\alpha}\,,
 \end{array}
\end{equation}
where $u^a(\phi)$ is a unit tangent vector (w.r.t. the kinematic metric $g_{ab}$) at $Q^a$ that is determined by the direction $\phi^A$ and $\kappa$ is the natural generalisation to the quantum realm of the classical $\kappa$ in \eqref{NbodyzeroE}, with $L$ being the scale variable appearing in a potential homogeneous in $L$ of degree $\alpha$, $V(L,q)=L^{\alpha}V_T (q)$, which is required to match the dynamics of the classical $N$-body system.\footnote{The appearance of a scale variable by no means fails to make the description relational. In fact, it is $\kappa$, an \emph{intrinsic} geometric property of the curve in shape space (related to its curvature radius), which becomes a degree of freedom entering the associated equation of state.} Finally, the function $\tilde{K}$ is also specified when reproducing the correct classical limit (see \citealp[][\S 3.3]{747}).

\subsection{Emergence of Born Statistics}

Two important remarks are in order at this point. Firstly, the second term in the equation for $\kappa$ in \eqref{debohm}---which is required to guarantee the correct classical limit---breaks the guidance equation, namely the equality between the shape momenta $p_a=\sqrt{\kappa}u_a$ and $\nabla _a S$ under evolution, thereby seemingly threatening the recovery of Born statistics. It is very important to emphasize that this poses no issues, given that the guidance principle is needed to yield the statistical predictions of QM for \emph{subsystems}. Hence, there is little meaning in recovering the Born statistics for the wave function of the whole universe. There is simply a unique global dynamics given by~\eqref{debohm} that, in principle, includes large-scale gravitational subsystems effectively described by Newtonian mechanics, along with small-scale quantum subsystems effectively described by QM. This single dynamics for the whole universe includes both standard classical and quantum theories. Thus, we simply need to analyze the dynamics of subsystems to assess the status of the Born rule. As discussed at length in \citet[][\S\S 2.2, 3.4]{747}, our approach will be to consider the formation of effectively isolated and dynamically decoupled bounded subsystems, which will then follow their own dynamical system, structurally the same as the global system~\eqref{debohm}. The appropriate regime may be described by the following three conditions:

\begin{enumerate}
\item The classical component of the total potential acting on a particular subsystem $S_I$ is dominant and entirely characterized by degrees of freedom within $S_I$ itself.
\item The quantum potential from the rest of the universe has a negligible effect on $S_I$. %This is essentially the expected effect of decoherence.
\item We assume either a temporal regime in which $\tfrac{\delta L_I}{L_I}\ll 1$, or, more generally, $\delta L_I \approx 0$, i.e., bounded subsystems have formed that asymptotically tend to a stable state with constant size.
\end{enumerate}

In such a regime, we effectively end up having decoupled dynamical systems for subsystems,  similar to~\eqref{debohm}, but, crucially, with no additional term for the equation of $\kappa$, because the relative size of these subsystems does not change ($\tilde{K}\propto\tfrac{dL}{L}$; see \citealp[][\S3.3]{747}). This will ensure that the effective wave function for a given subsystem guides its evolution through standard de Broglie-Bohm dynamics, which will, in turn, lead to equivariance, and hence, consistency of the Born rule for a probabilistic distribution for this subsystem.

Secondly, there is an important conceptual difference with the non-relational theory that should be stressed: Whereas the latter considers an ensemble of universes with a distribution given by the quantum equilibrium condition, we assume that the relational system describes the evolution of a single universe. This is because, as discussed in the following section, we regard a solution of \eqref{debohm} (i.e., a single curve in shape space) as an individual possible world---not an ensemble thereof. This is a purely interpretive move that does not alter the physical content of the model: We just regard the notion of ``ensemble of worlds'' as epistemic (i.e., a mental construction) and not as a literal collection of possible worlds. Indeed, it is a well-known fact that an epistemic ensemble of universes, once restricted to one element, gives the same probability distributions for the associated subsystems as a real, single universe does for its subsystems---provided the subsystems in both cases are effectively isolated and similar \citep{252}. 

\subsection{Quantum Rods and Clocks}

Remarkably enough, the dynamics exhibited by the equation of state \eqref{debohm} shows the attractor-driven behaviour in shape space already stressed in the classical case (\citealp{706}, \citealp[][\S\S3.5-3.6]{726}). In a nutshell, this means that the direction of secular growth of complexity---which, recall, measures the amount of structure or ``clustering'' inherent in a shape--- \emph{defines} the arrow of time (see \citealp[][\S 3.5]{747}, for the numerical analysis of the 3-body system, which highlights how the attractor-driven behaviour in shape space is independent of Planck's constant). As is well known in the literature, the asymptotic evolution of the classical $N$-body system gives rise to the formation of stable substructures, among which the so-called \emph{Kepler pairs} play a major role as physical rods and clocks. Clearly, a full technical characterization of the quantum counterpart of these Kepler pairs is part of the subtle question of the formation of subsystems satisfying effective Schr\"odinger equations in the quantum model. The case of the 3-body system is certainly a promising result towards this end. In this regard, let us once again emphasize that the decidedly holistic character of our model is inherited from the classical theory, whereby subsystems march in step asymptotically. Once extended to the quantum realm, this holistic approach naturally accommodates entanglement, which otherwise holds in our model essentially as in the standard case.  

\subsection{Classical Limit}

The de Broglie-Bohm approach allows a simple analysis of the classical limit: Whenever the evolution of the quantum potential $V_q=-\frac{\Delta\,R(q)}{2\,R(q)}$ along the curve becomes negligible, the evolution of the curve in shape space is effectively described by classical equations of motion determined by the full potential $V_T=C+V_q$. Given the nuanced status of the wave function in this model (see \S\ref{sec:wf}), the understanding of the exact physical conditions that guarantee that the evolution of the quantum potential is negligible---with a relational counterpart of the standard decoherence mechanism being possibly involved---is still a subject of investigation. For the purpose of the philosophical analysis carried out in this paper, however, it is not necessary to point out what these conditions are: It is sufficient that the classical regime is in fact attained when $V_q$ goes to zero. Indeed, in this case, \eqref{debohm} reduces to the equation of state associated with the classical $N$-body system (\citealp[][\S 3.3]{747}):

\begin{equation}\label{new}
 \begin{array}{rcl}
  d q^a &=& u^a(\phi)\,,\\
  d \phi_A &=& \frac{\partial \Phi_A}{\partial q^a}u^a(\phi)-\frac{\partial \Phi_A}{\partial u^a} \left(\frac 1 2 g^{bc}_{,a}(q)u_b(\phi)u_c(\phi)
  +\frac 1 \kappa C_{,a}(q)\right)\,,\\
  d \kappa &=& -2u^a(\phi)C_{,a}(q)\mp\gamma\kappa\sqrt{-\left(1+\frac{2\,C(q)}{\kappa}\right)}\,,\\
 \end{array}
\end{equation}
where $\kappa := p^2/L^{\gamma}$, with $L^{\gamma}$ being the scale component of the classical potential, $V(L,q)=L^{\gamma}C(q)$. In order for \eqref{debohm} to reduce to \eqref{new}, the homogeneity degrees of the classical and quantum potentials must match, $\gamma =\alpha$ (recall discussion following \eqref{ukappadefinition}). As expected, \eqref{new} is nothing but \eqref{NbodyzeroE} (the homogeneity degree $k+2$ has been redefined as $\gamma$, for simplicity).

\subsection{Comparison with Similar Proposals}

In the recent literature, another attempt has been made to define the de Broglie-Bohm dynamics on shape space \citep{530}. Although on the surface our model is similar to the D\"urr, Goldstein, and Zangh\`i's one in that both attempt to provide a relational account of quantum dynamics within the general framework of SD, a number of crucial differences between these authors' approach and ours definitely remain. First and foremost, in our view these authors do not embrace the fundamental insight of SD, namely, that shape space is thought of as the fundamental arena where \emph{all} physics unfolds. According to this insight, laboratory physics is arrived at from the holistic fundamental dynamics in suitably effective regimes \emph{within} shape space. These effective regimes allow us to write shape space as the Cartesian product of configuration spaces of subsystems, thereby effectively expressing the fundamental dynamics in terms of separate dynamical subsystems in each subspace. In particular, local frames of reference, defined through distant stars and ephemeris time, are local shape substructures within the global shape of the universe. This is in stark contrast with the spirit of \cite{530}, where shape space seems to be taken as a convenient mathematical structure wherein physics is simplest (whence their insistence on geodesic dynamics), only to get back standard configuration space physics through suitable liftings and gauge fixings within their fibre bundle structure. 

Moreover, the approach of \cite{530} implies that the wave function is non-normalizable precisely due to their insistence upon describing the interesting physics in configuration space, not shape space. Unlike theirs, ours is a dynamics that always unfolds in shape space, as emphasized above. This comes with the bonus that the shape space of the $N$-body system, the shape sphere, is \emph{compact}, which guarantees the normalizability of the wave function. This is certainly desirable if one is to endorse the familiar interpretation of $|\Psi^2|$ as a probability distribution. The authors of \cite{530} argue that the non-normalizability arises from unphysical differences and, further, that it is the universal wave function that fails to be normalizable, while there being no reason why the conditional wave functions, the ones accounting for laboratory physics, should fail to be normalizable. This may be true, but the issue is clouded by their approach, whereas in ours the matter is readily seen.

Most importantly, these authors have not taken into account the recent developments in SD brought forth by the insight in \cite{706} that growth of complexity defines the arrow of time. Although their unparametrized curve in shape space and associated notion of relational time do certainly share some of the early insights of SD we favor, their model is not able to address the emergence of an arrow of time and subsystems formation, which \emph{are} obtained in our approach thanks to the crucial attractor-driven behaviour of complexity in shape space. Indeed, given that D\"urr, Goldstein, and Zangh\`i construct a geodesic theory on a \emph{compact} configuration space---which is the case of $\mathcal S$ in the $N$-body system---, their model exhibits Poincar\'e's recurrence and, hence, is unable to accommodate stable structure formation. Unlike this, the crucial role of $\kappa$ in our model, yielding a modified geodesic dynamics, enables us to avoid  Poincar\'e's recurrence.

The previous points lead us straight to our next remark, namely, that the quantum model presented in \cite{530} follows the principles of standard de Broglie-Bohm theory much closer than it does with the principles of Barbour-Bertotti theories. As a result, the D\"urr, Goldstein, and Zangh\`i's model exhibits a well-defined guidance equation \emph{by construction}, hence being able to reproduce the predictions of quantum mechanics just as standard, non-relational de Broglie-Bohm theory does. However, the approach falls short of providing a concrete step towards a fully relational quantum theory, in a Leibnizian/Machian spirit: It is more of a mathematical play---a ``shape space rendition'' of the standard theory--- rather than an attempt at a new approach to quantum physics. 

To the contrary, our de Broglie-Bohm model follows the spirit of SD to the core, with the intent to open a window into new physics. Indeed, from a universal perspective, our model is substantially different from the standard de Broglie-Bohm theory. Recall that \eqref{debohm} does not automatically encode the guiding principle and the ensuing Born rule interpretation of $|\Psi|^2$, which are recovered only on a subsystem perspective. This perfectly fits the Leibnizian/Machian spirit of the framework in that these aspects of the physical description are invoked only when they do have observable consequences. Thus, metaphysically perplexing questions like ``What is the meaning of the modulus square of the universal wave function?'' are avoided \emph{ab initio}.

Finally, it is worth stressing that the departure of our model from the standard de Broglie-Bohm theory has no major implications for \emph{what} the theory is about. In other words, our model depicts the dynamical development of the couple $(Q^a,\Psi(q))$, which is just the natural relational counterpart of the subjects of dynamical development in the standard theory. Hence, the departure simply amounts to a richer dynamics, but it does not impact substantially on the metaphysics of the theory---to which we now turn.

\section{Self-Subsisting Structures in a de Broglie-Bohm Setting}\label{subsec:nopsi}

The key point when supplying an interpretation of the model \eqref{debohm} is to recall that the quotienting out operation that leads from standard configurations to shapes represents a structural reconceptualization of the starting non-relational physical description: What in the starting picture is intended as a collection of individual entities endowed with intrinsic properties, is now a mosaic of relations among identityless and propertyless relation-bearers. In this sense, PSD's theoretical machinery already carries some minimal yet defining metaphysical traits even before a full metaphysical characterization is supplied. The added value of such an approach is that it dissolves by construction any issue related to the ontological underdetermination inherent in physical theories. In the present case, it does not matter whether the ``original'' de Broglie-Bohm theory is intended to fundamentally describe, say, $N$ material particles moving in a $3$-space, or a world-particle moving in a $3N$-space: The quotienting out procedure, in the end, will preserve all and only those structural aspects of these two interpretations that are needed to back up empirical observations. Given that the two interpretations are empirically equivalent, these preserved structural aspects will be exactly those common to them.

Even though the PSD formalism already comes equipped with some metaphysical baggage, this is not to say that a full-fledged metaphysics is univocally picked by this theoretical framework. To the contrary, PSD's parsimonious commitments are compatible with a surprisingly wide range of metaphysical tastes (on this score, see \citealp[][\S4.3]{vasnarkos1}). In the present case, two choices seem the most straightforward.

On the one hand, one may take the shape space picture at face value, and maintain that the wave function is a field defined over this space, which determines the dynamical behavior of the de Broglie-Bohm shapes as encoded in an unparametrized curve representing a solution to \eqref{debohm}. This view can be coupled with some sort of monism \emph{\`a la} \citet{538}, according to which a shape is an integrated whole that grounds the existence of its parts (i.e., relations and \emph{relata}), thus constituting a ``world-particle.'' This position would be some sort of relational counterpart of David Albert's ``marvelous point'' view:

\begin{quote}
On Bohm's theory [\dots] the world will consist of exactly two
physical objects. One of those is the universal wave function and the other is the universal particle. And the story of the world consists, in its entirety, of a continuous succession of changes of the \emph{shape} of the former and a continuous succession of changes in the \emph{position} of the latter.\\
\citep[][p.~278, his emphasis]{376}
\end{quote}

Albert's quote highlights how the picture he advocates requires some non-trivial structure to make sense of the wave function's and the world-particle's dynamics. First, the universal particle is \emph{located} somewhere, while the wave function possesses a \emph{shape}. Clearly, this means that both objects exist in, and are characterized by means of, an external embedding space. Second, both wave function's shape and world-particle's position \emph{change}, which means that some external time-like ordering has to be postulated as a reference for labeling the different dynamical stages. Now, in the standard de Broglie-Bohm theory, there is nothing problematic in postulating this extra structure. After all, the theory's dynamics is quite manifestly referred to spatiotemporal background structures: Albert's interpretation just assumes these structures to be a $3N$-dimensional space and a continuously flowing universal time. Things, however, become less tenable when trying to transpose this picture to the shape space setting in PSD. Indeed, the whole point of the mathematical reduction of the standard de Broglie-Bohm dynamics to \eqref{debohm} is exactly to get rid of the background structures on which the non-relational dynamics relies. This seems to suggest that shape space should be interpreted as a (structured) collection of relational configurations rather than a literal embedding space in which such configurations are placed. A further clue that goes in this direction is that the kinematic metric defined on $\mathcal{S}$ does not measure any spatial distance (irrespective of how loose we take this concept to be) but just how (dis-)similar shapes are. Even a declared shape space realist like Julian Barbour himself regards this space as an embodiment of the fact that all the possible relational configurations exist all at once, timelessly \citep{10}. This is why \citet[][\S3.1.3]{402}, regards Barbour's position as akin to David Lewis' modal realism. Note how this move is not available to a de Broglie-Bohm shape space realist, given that the theory privileges the description of a single system---the \emph{actual} one---not an ensemble of possible copies of it. In light of this discussion, accepting the elimination of external embedding structures that leads to \eqref{debohm} only to then postulate shape space as the place where the dynamics takes place means kicking the background structures out of the front door just to let them sneak in again through the backdoor.

On the other hand, another readily available interpretation of \eqref{debohm} consists in fully embracing the structural spirit of PSD, and conceiving of de Broglie-Bohm shapes as structures in a genuine ontological sense. By itself, this choice does not mark a huge difference from the previous position. After all, also the world-particle realist would have no problem in accepting that such a particle can be decomposed in smaller ``pieces.'' However, she would claim that this fine-grained picture is not fundamental but grounded in the world-particle as an integrated whole. Thus, the ontic structural realist has to make some extra steps in characterizing her position. The first step is, obviously, to turn the monist picture upside-down, thus claiming that the concept of shape supervenes by way of composition on a fundamental mosaic of relations and \emph{relata} that are equally fundamental and mutually dependent ontologically. Such a moderate take on structures is preferable to an eliminativist one (according to which only relations are fundamental) because it makes it easier to argue that shapes are \emph{concrete} physical structures as opposed to abstract constructs (see \citealp[][\S2]{702}, for a fully worked-out argument for preferring moderate ontic structural realism over its eliminativist cousin). This is especially important if we want to argue that relational parts literally add up to make a shape. 

All of this is obviously not enough, as still the question remains open as to ``where'' the shape is placed. In order to reject from the get-go any answer to this question that suggests a realist commitment towards shape space (either in a substantivalist or a modal realist fashion), it is sufficient to shift our commitment to dynamical curves. Invoking this move is anything but controversial if we think that a solution of \eqref{debohm} (and, in general, of \eqref{curve0}) represents the entire history of relational evolution at a universal level and, thus, encodes all the fundamental physical information that, according to PSD, the entire history of the universe can contain. Consequently, a dynamical curve can be interpreted as a possible world. One may wonder whether this is just a more convoluted way to state Barbour's commitment to shape space as the collection of all possible relational configurations: After all, the collection of dynamical curves is likely to span $\mathcal{S}$ in its entirety. This objection can be repelled on a twofold ground. Firstly, whereas a full commitment to shape space does not postulate any particular ordering of the relational configurations---besides the differential structure encoded in the kinematic metric---, a commitment to dynamical curves as possible worlds places a stricter constraint on the modal horizon of the theory. Indeed, there are shapes that belong to some possible worlds but not to others: The ``modal cauldron'' of shape space is divided into self-contained and mostly non-overlapping modal units---i.e., possible worlds. Secondly, and most importantly, the curve-as-possible-world commitment need not be realist with respect to such \emph{possibilia}: There is only one of such worlds that deserves a full realist commitment, i.e., the \emph{actual} one. Mathematically, this world is represented by the curve generated by the initial conditions that obtain at our world. 

Having established how to individuate worlds, the next question is what it is that makes up a world (possible or actual). This question can be answered by considering that an unparametrized curve is nothing but a succession of shapes related by a weak topological ordering. In other words, there is just a shape after another after another, and so on, without any fundamental fact about how ``far'' they are from each other or in which direction the ordering goes. This ordering is certainly too weak to carry any substantial temporal connotation. Therefore, an unparametrized curve can be taken as a representation of the fact that, in a possible world according to PSD, all relational configurations are given all at once in a timeless sense. As argued in the companion paper, this machinery is enough to constitute a sufficient Humean supervenience basis not only for spatial and temporal concepts---such as scale and duration---but also for dynamical parameters such as mass, charge, spin, and the like. In conclusion, we submit that the metaphysics of self-subsisting structures we propose represents a good fit also for a non-relativistic de Broglie-Bohm setting.

\section{Self-Subsisting Structures and Wave Function}\label{sec:wf}

According to our metaphysical framework, a possible world governed by the relational version of the de Broglie-Bohm laws does not feature a Humean mosaic very different from a world governed by the classical laws \eqref{new}. However, the two worlds radically differ in the way these structures are dynamically ordered. This is an evident consequence of the fact that the relational de Broglie-Bohm equations introduce many more degrees of freedom than \eqref{new}: Indeed, the way the classical picture \eqref{new} is recovered from \eqref{debohm} consists in damping these extra degrees of freedom by letting the quantum potential go to zero. Physically, this means that the ``emergent'' particle motions are much more complicated in the quantum world than in the classical one. This added complexity accounts for the non-local nature of the particles' motions (hence accounting for the empirically tested violation of Bell's inequalities).

By inspecting \eqref{debohm}, it results clear that some of the degrees of freedom describing the dynamics are not strictly speaking inherent in the dynamical curve---i.e., they are not geometric properties of said curve. We are of course referring to $R(q)$ and $S(q)$, which are the wave function's amplitude and phase. In other words, there is something not geometrically encoded in the sequence of configurations making up the dynamical curve that, nonetheless, contributes to the description of the behavior of the system. This poses a challenge to our metaphysics, since it seems that a structure external to the system---the wave function---is needed in order to make sense of the dynamical development. The obvious solution to this problem is to claim that the wave function does not refer to any element of the fundamental ontology over and above self-subsisting structures---it is just a further codification of what self-subsisting structures are. Such a solution can be achieved in many ways. In the following, we will analyze the broad lines of research. 

Let's start from the nomological interpretation of the wave function. In the literature, some extensive attempts have been made to defend the view that the wave function in a de Broglie-Bohm setting has not to be intended as a \emph{sui generis} physical object. Instead, it should be regarded as a law-like element of the formalism, akin to the Hamiltonian function in classical mechanics (cf. \citealp{428}, \S 4.1., for an articulation of the argument). This reasoning has been pushed further in a Super-Humean context by claiming that the wave function in the de Broglie-Bohm theory supervenes on the entire mosaic made up of permanent particles related by ever-changing distance relations (see, e.g., \citealp{485}, \S3.2). Following this line of reasoning, $R(q)$ and $S(q)$ would have to be intended as mathematical artifacts that ``live outside'' self-subsisting structures, but in fact referring to intrinsic features of such structures. As such, the wave function should be regarded as supervening on the fundamental mosaic as a descriptive layer in the same way temporal notions do. Although this solution goes in the direction of ontological parsimony, we are reluctant to fully embrace it for three main reasons.

First of all, it is doubtful whether the form of $\Psi$ is determined in any cogent technical sense by the underlying particles' dynamics (be it standard or relational). For example, as pointed out in \citet[][section 2.3]{687}, while the classical Hamiltonian can be constructed as a function of particles' positions and velocities (thus lending support to the view that it supervenes on the mosaic of particles), such a story is not readily available for the wave function. Indeed, this latter function is just a solution of the Schr\"odinger equation, whose specification does not need direct information about the particles' positions and velocities (or the relational configurations, in the present context). This, in turn, suggests that there is more to the wave function than just its being a compact way to describe the particles' change in position over time (or the change in their relational configuration). This is, of course, not enough to knock down the nomological take on the wave function but is enough to highlight the fact that it is controversial whether the formalism of the de Broglie-Bohm theory (both in its standard and its relational versions) really lends support to the claim that the wave function nicely and simply supervenes on more fundamental features of reality. Secondly, in the case of self-subsisting structures, it would be hasty to consider the wave function as supervenient on the mosaic represented by a dynamical curve in a way similar to how temporal notions, such as time's arrow, supervene on such a mosaic. In fact, there is a clear dissimilarity between supervenient temporal notions and $\Psi$: While the former have no mathematical representation living directly on shape space---and, in fact, appear only when the ephemeris constructions are carried out \citep[][\S4.2]{vasnarkos2}---this is obviously not the case for the wave function. Thirdly, in the companion paper (\emph{ibid.}, \S4.1), we made it clear that our ontic commitments were aimed at finding the minimal amount of facts needed to make sense of all and only the degrees of freedom that are common to all models of \eqref{curve0} (i.e., the couple $\langle q^a, \phi^A \rangle$). In the present case, it is quite obvious that $R(q)$ and $S(q)$ cannot be written in terms of $\langle q^a, \phi^A \rangle$ \emph{only} so, according to our norm for ontic commitments, $\Psi$ cannot be said to represent just fundamental facts about self-subsisting structures.

Another possible solution to the problem may be to find a technical way to ``absorb'' $R(q)$ and $S(q)$ in the geometric degrees of freedom of the curve. Unfortunately, this option does not seem viable within the model \eqref{debohm}. Indeed, the dynamical system \eqref{debohm} contains second derivatives of both $R(q)$ and $S(q)$, which cannot be encoded in the geometric degrees of freedom ($\phi_A \,,\kappa$), as these contain information only up to first derivatives (cf. \eqref{ukappadefinition}). This mathematical fact shatters any hope of regarding the geometrical properties of a single curve in shape space as sufficient to match all the predictions of quantum mechanics---some additional information external to the curve has to be provided. A possible way out may be the following: Instead of considering a single de Broglie-Bohm curve, consider the set of curves generated by the set of all possible boundary conditions. For each shape $q_i^a \in\mathcal S$, with $i$ labelling the set of possible curves, consider the curve $q_i^a(s)$ associated with the boundary conditions ($\phi_i^A\,,\kappa_i$). This way, the second derivatives of both $S(q)$ and $R(q)$ can be expressed in terms of ($\phi_i^A\,,\kappa_i$), whereby the phase is obtained up to an unobservable global phase and the amplitude up to normalization. Unfortunately, even if such a mathematical reduction of the wave function is in principle implementable, it would effectively amount to taking an exceedingly large number of degrees of freedom, thereby making the dynamical system essentially intractable (see \citealp{747}, for details). The good news is that PSD, as a general framework, does not rule out \emph{by construction} the possibility of mathematically reducing $\Psi$ to the geometric degrees of freedom of, at least, a set of curves. However, it may be the case that such a reduction would result more natural (and tractable) in quantum models different from \eqref{debohm}. This is a line of research currently in progress (see section \ref{sec:disc} for some remarks on this score).

There is perhaps a more viable way out of this impasse, which requires however a bit of justification. Let's start with the simple observation that the classical dynamics \eqref{new} is arrived at by letting the quantum degrees of freedom in \eqref{debohm} ``die off,'' as represented by the quantum potential going to zero. So, the solutions of \eqref{new} may be seen as particular cases of solutions of \eqref{debohm} where quantum effects are so much damped to be absent. This encourages us to look at classical PSD models as some sort of \emph{degenerate} quantum models, i.e., models where the quantum degrees of freedom encoded in $R(q)$ and $S(q)$ are ``inactive'' and, thus, modelled by an identically null wave function. On top of this, we can add that quantum degrees of freedom have a peculiar nature in the PSD framework in that they are not geometric degrees of freedom, so it does not make much sense to consider them as non-fundamental just because they are not part of the couple $\langle q^a, \phi^A \rangle$ (as, instead, is the case of other geometric degrees of freedom like the curvature $\kappa$). If we accept all of this, then we can say that it is not much of a twist of our initial ontic commitments if we revise our norm to include quantum degrees of freedom in the fundamental picture of reality. So, we can restate such norm as: ``We ought to include in the fundamental ontology of the world only the minimal amount of facts needed to make sense of all and only the degrees of freedom that are common to all models of \eqref{curve0}, namely, $\langle q^a, \phi^A, R(q), S(q) \rangle$.'' Note how such a revision of our ontic commitments comes from a deeper investigation of the solution space of \eqref{curve0}, thus showing how our proposed metaphysics is susceptible to changes in the physical understanding of the world and not rigidly fixed \emph{a priori}.

This revision of our ontic norm clearly bears substantial conceptual consequences on our metaphysical framework. We immediately see this when attempting to answer the question about what it is that $R(q)$ and $S(q)$ refer to. For starters, they cannot refer to something beyond shapes, as this would violate the spirit of PSD as an intrinsic description of a physical system. Secondly, they should not refer to intrinsic properties of the \emph{relata} in a self-subsisting structure, as this metaphysics dispenses with fundamental monadic properties. Instead, our answer is very simple. $R(q)$ and $S(q)$ do not refer to anything new in the ontology; they just represent a refinement of the characterization of what a self-subsisting structure is. In other words, $R(q)$ and $S(q)$ tell us that the relations making up such a structure are not \emph{just} spatial, but contain also a ``quantum component.'' To put it more explicitly, self-subsisting structures in a de Broglie-Bohm setting are held together by hybrid spatial/entanglement relations. A detailed account of how this ``metaphysical hybridization'' works deserves a paper on its own. Suffice it to say that, in this particular case, hybridization means that (i) the \emph{relata} are fully individuated only when both the spatial and the entanglement ``components'' of the relation are specified, and (ii) there is no real ontological difference between the two ``components.'' In short, there is just one type of relation at the fundamental ontological level, not the combination of two distinct ones. This begs the question as to why, historically, people have developed two distinct accounts of spatial and entanglement relations. The answer is that, in everyday life, we have the impression that spatial and entanglement relations are distinct entities because we are acquainted with the ``classical side'' of the world, where the quantum aspects of these relations are hidden from view, while we tend to look at quantum phenomena as an added layer of reality that manifests itself in the appropriate regimes. Once we realize that the whole universe is quantum, all the pieces fall in place. The suggestion that fundamental physics reveals the existence of a host of hybrid relations has already been made in the literature, especially in the context of general relativity (cf. \citealp{582,557}).

It is quite obvious why we prefer a \emph{unique} hybrid relation instead of two \emph{separate} ones: In this latter case, we would have two distinct types of structures as well---so that a self-subsisting structure would be some sort of ``superposition'' of spatial and entanglement structures. This would introduce a confusing dualism in the ontology where, in fact, there is no need for it. Indeed, spatial and entanglement relations can be seen as perfectly overlapping, in the sense that they take in the same \emph{relata} and induce the same order among them. To see this, it is sufficient to point out that, in a de Broglie-Bohm setting, both these relations are irreflexive, symmetric, and connected. This means that there is no pair of \emph{relata} standing in a spatial relation without also standing in an entanglement relation: The hybrid relation we are positing can be seen as some sort of ``relational merger'' of the two. One may object that it is an empirical fact that there are non-entangled systems that nonetheless stand in spatial relations, but this would confuse the global perspective with the subsystem one. It is true that, whenever the right conditions obtain for the definition of an effective wave function, we can define an associated subsystem that behaves \emph{as if} it was isolated from the rest of the universe. However, this perspective is just useful for practical purposes---e.g., the description of measurements in a laboratory setting---but it does not reflect the fundamental reality, i.e., that all particles in a de Broglie-Bohm setting are entangled together.

The hybrid nature of the relation making up a de Broglie-Bohm self-subsisting structure is consistent with our view that the quotienting out procedure carried out on the starting non-relational model \emph{determines} the relational features of these structures. Indeed, the spatial and quantum degrees of freedom encoded in the hybrid relation are what remains of the external embedding space \emph{and} the non-relational universal wave function after the quotienting out operation is performed.

Adding an entanglement dimension to the relations making up the mosaic of self-subsisting structures---represented by a solution of \eqref{debohm}---does not undermine the world-building character of such a relation. It is easy to see that this relation is perfectly natural (that is, it is not disjunctive), it is pervasive in that it ``chains together'' all the places in a structure, it individuates unambiguously all the \emph{relata}, and it does not supervene on the intrinsic natures of these \emph{relata} considered separately---i.e., it is an \emph{external} relation. In short, the spatial/entanglement hybrid relation qualifies as a world-building relation by the standards set out by David Lewis himself \citep[cf.][pp.~75--76]{284}. The choice of entanglement as a world-building relation for the quantum regime---which supplements or even supplants spatiotemporal relations altogether---has already surfaced in the literature (see, e.g., \citealp{579,entl,jak}), but introducing entanglement in the context of PSD renders the proposal far less controversial. Suffice it to mention that the objection according to which entanglement structures cannot be considered physical by themselves ``unless they are implemented or instantiated in spacetime'' \citep[][p.~6]{293} is quite easily addressed in the present context: There are no entanglement structures \emph{simpliciter} but hybrid structures held together by relations that possess a spatial connotation besides the quantum one.

There are four objections that surface at this point. The first is that the entanglement relation, as empirically exemplified by EPR correlations, is \emph{independent} of spatial distance. The statistical correlations observed in a pair of electrons in, say, a singlet state do not change if we repeat the measurement on identical systems whose parts are placed at different spatial distances. How can this be described by a \emph{single} relation holding between the electrons? Such an objection doubly misses the mark, in our opinion. First, positing a relation with a spatial as well as an entanglement dimension does not imply that one of such dimensions should be dependent on the other. If that was the case, one of the dimensions could be functionally reduced to the other, thus collapsing the relation to a ``pure'' one. Secondly, the objection confuses types with tokens. In the previous example, a singlet system whose component electrons are, say, 5 meters apart and a singlet system whose electrons are 500 meters apart represent two distinct exemplifications of the same hybrid relation type. Now, if we focus only on the entanglement part of these occurrences, they are indistinguishable, but, if we add the spatial dimension, the two systems are perfectly distinguishable, as we expect them to be in a de Broglie-Bohm setting, where the spatial characterization of the particles is an integral part of the physical description of the systems. 

This last consideration is relevant also for the next objection, which can be articulated as follows. This hybrid relation you postulated is an even worse case of metaphysical gerrymandering than just taking it as the disjunction of two distinct relations. In fact, you are taking two concepts that are physically unrelated and merging them by \emph{fiat}. What point of contact could there be between, say, ``X and Y being in a singlet state'' and ``X and Y being $n$ meters apart''? This objection overlooks the spirit of the de Broglie-Bohm theory, as we see it. For us, the focus of the theory is quite clearly to recast non-relativistic quantum mechanics as a theory of material particles, all ``feeling'' each other via universal entanglement and, \emph{as a consequence}, moving in a choreographed way. On this view, entanglement and spatial relations are inextricably connected from a physical standpoint: Particles do not just change their positions over time, they do so \emph{in step}. This added dimension to their motion is given by the quantum degrees of freedom encoded in the wave function. The PSD model \eqref{debohm} just translates this view into purely relational terms. But even beyond the context of the de Broglie-Bohm theory, many hypotheses have recently surfaced in the physical literature, suggesting that space and entanglement may be deeply intertwined concepts \citep{entspa2,entspa3,entspa}.

The third objection is that, by adding such an entanglement dimension to the relation making up self-subsisting structures, we are smuggling some irreducibly modal features in the fundamental mosaic, \emph{contra} the Humean spirit of our metaphysics. Such an objection can be cast in terms of the singlet electron system in the following way: The holding of the relation of anticorrelation between electrons in the singlet state amounts to there being a necessary connection between the values of the spin of both electrons, in the sense that fixing the value of the spin of one electron automatically determines the spin of the other. To this objection, we reply that the necessary connection depicted by the example does not hold between the electrons, but between their spins. But, according to our metaphysics, there is no fundamental property called ``spin'' in the mosaic. In other words, the objection confuses a fundamental fact about a self-subsisting structure (a certain spatial/entanglement relation among two \emph{relata}) with an ``augmented'' description of the fact in terms of a two-electron system in a singlet spin state. As in the case of temporal concepts, such a description uses vocabulary shortcuts, such as ``electron'' and ``spin,''
to refer to the fundamental underlying fact, but this does not imply that individual entities like electrons and intrinsic properties like spin are part of the ontology. In this sense, the type of necessity involved is more conceptual than metaphysical. Indeed, a statement like ``The spin of electron $1$ is anticorrelated with the spin of electron $2$'' can be seen as a (conventional) definition of the higher-order language terms used to describe the underlying fundamental fact. If we accept this, then claiming that fixing the value of the first spin necessarily fixes the value of the second is tantamount to claiming that Hector's death automatically renders Andromache a widow---the ``status change'' is due to how the terms are defined, but does not represent the change in any fundamental fact.

Finally, it may be pointed out that adding an entanglement nature to the external relations making up a self-subsisting structure does, again, harm to the Humean spirit of our metaphysics because it adds a holistic component that cannot be reconciled with the local nature of the Humean mosaic, which should instead consist of just ``one thing after another.'' If this objection is meant to highlight that the complete physical information regarding a part of a self-subsisting structure depends on the shape as a whole, then we totally agree---this is at the root of the Machian spirit of PSD, as exemplified in the passage quoted in the introduction. From this point of view, our metaphysics has a holistic character already at the classical level. Does this betray Humeanism? We would say no, for the simple fact that the spatial component of the world-building relation making up self-subsisting structures supports a well-defined mereological structure distinguishing the whole from its fundamental constituents. In this sense, introducing entanglement in the picture does not impair the view that the complete physical characterization of a self-subsisting structure is determined by the ``pattern'' of its fundamental parts. Still, one may object that the physical characterization of a ``subregion'' of the structure (i.e., a composite subsystem) would fail to supervene on the components of that subregion only. This is certainly true but, in a de Broglie-Bohm setting, the notion of composite subsystem is just an \emph{approximation} used whenever the rest of the universal system can be safely neglected for practical purposes. For this reason, we do not accord any ontological meaning to anything beyond a whole structure and its fundamental constituents. From this point of view, our position is akin to the ``austere Bohumianism'' described in \citet[][\S5]{606}.

\section{Self-Subsisting Structures and Primitive Ontology}
\label{subsec:PO}

The de Broglie-Bohm theory is often associated with a general approach to quantum physics called \emph{primitive ontology} approach (see \citealp{203}, for a complete analysis of its defining features). Schematically, the primitive ontology approach to quantum physics consists in postulating (i) a fundamental set of material entities occupying a fixed spacetime background and (ii) a non-local dynamical law---formally depending on the quantum state---that describes the temporal development of this fundamental set of entities. The primitive ontology approach is usually mentioned in relation to the measurement problem of standard quantum physics, that is, the problem of reconciling the linear and unitary Schr\"ondiger evolution---which leads to quantum superpositions and interference---with the appearance of definite outcomes when actual measurements are performed on quantum systems (see \citealp{197}, for a thorough presentation of the issue). The primitive ontology approach is put forward as a solution to the measurement problem because it provides a clear physical description of measurement-like interactions in terms of matter dynamically evolving in spacetime. A determinate outcome of an experiment is then just due to matter ending up in a particular configuration at a certain time as a result of its dynamical development. In this sense, following \citet{415}, primitive ontology theories shift the accent from physical observables to \emph{local beables}, i.e., material stuff occupying space at a given time. Hence, the spirit of the primitive ontology approach is to provide a straightforward story that connects theory to observation. This point is made extremely clear in the following passage taken from \citet[][p.3159]{228}:

\begin{quote}
Our ability to reliably observe [empirical] facts is not itself derived from the physics: it is rather a presupposition used in testing the physics. So the contact between theory and evidence is made exactly at the point of some local beables: beables that are predictable according to the theory and intuitively observable as well. [...] Collections of atoms or regions of strong field or regions of high mass density, because they are local beables, can unproblematically be rock-shaped and move in reasonably precise trajectories. If the theory says that this is what rocks really \emph{are}, then we know how to translate the observable phenomena into the language of the theory, and so make contact with the theoretical predictions.
\end{quote}

The question is, then, whether the model \eqref{debohm} belongs to such an approach, given that it represents a quotiented out version of the de Broglie-Bohm theory. By looking at the characterization of the approach given above, it is easy to answer the question in the negative. For starters, \eqref{debohm} formally is a local law in shape space, \emph{contra} condition (ii). Now, of course, also the equations of the standard de Broglie-Bohm theory are local in configuration space---this is the starting point of Albert's marvelous point theory, after all. However, the proponents of the primitive ontology approach would insist that the ``local'' guiding equation $\boldsymbol{\dot{Q}}=\nabla S$ is just a compact way to write down $N$ coupled equations $\boldsymbol{\dot{Q}}_{i}=\nabla S$, one for each particle ($\textbf{Q}=\langle\textbf{Q}_{1},\dots,\textbf{Q}_{N}\rangle$). This way, they can translate the particle dynamics from configuration space to ordinary space. A similar move is obviously not available in the PSD case, shape space being the only ``arena'' for the unfolding of the dynamics. A dynamical description of $N$ particles moving in ordinary space can indeed be achieved at a certain stage, but it would be just an ``embellished'' description obtained by adding non-physical degrees of freedom to the picture---which means that it would carry no ontological weight. Moreover, \emph{contra} (i), PSD is not a theory of matter \emph{in} spacetime in any physical domain, quantum or otherwise. To sum up, if the primitive ontology approach identifies the fundamental ontology of the world with Bell's local beables, then the PSD framework is not faithful to such an approach.

However, it is already well-established in the literature that condition (i) has to be relaxed in some way for the primitive ontology approach to stand any chance of being extended to the quantum-gravitational regime, where dynamical geometry plays an important role. More precisely, to extend the approach to a (quantum) general relativistic context, it is necessary to let go of any fixed background spatiotemporal structure. Instead, the approach should include \emph{all} spatiotemporal degrees of freedom in the primitive ontology itself (see, e.g., \citealp{433}, for a discussion of background independence in the primitive ontology context, and \citealp{161}, for a concrete model that treats non-material degrees of freedom as part of the primitive ontology). If condition (i) is so weakened to accommodate background independence, then the PSD framework becomes more akin to a primitive ontology approach. This is trivial because, according to the metaphysics we propose, PSD is a theory of material structures whose characterization and dynamics are independent of any external spatiotemporal structure.

Note, however, that it is not that straightforward that softening condition (i) would be compatible with the original spirit of the primitive ontology approach. Indeed, if the notion of primitive ontology is divorced from that of background spacetime, then its identification with local beables is put in severe jeopardy since the latter is just material stuff occupying a fixed spacetime region. Suppose that the primitive ontology is no longer made of local beables in the strong sense used above. In that case, the whole approach loses its traction because no intuitive story in terms of spatiotemporal composition can be invoked to establish a link between theory and observation (e.g., measuring apparatuses being literally made of particles). Such a problem does not impact PSD since (a) it is a theory of self-subsisting structures, which are not strictly speaking local beables, and (b) it \emph{does} provide an intuitive story for how to get to local beables---which involves the dynamical mechanism that leads to stable subsystems formation, and the ephemeris constructions discussed in the companion paper.

Setting quantum gravity aside, it may \emph{prima facie} seem that a primitive ontology reading of the de Broglie-Bohm theory fares better than the PSD relational account as far as the reconstruction of the manifest image of the world is concerned. However, is it really the case? Otherwise said, is it the case that macroscopic objects in standard de Broglie-Bohm theory are just conglomerates of Bohmian particles? A moment of reflection shows that such a claim is, at best, very optimistic. Indeed, Bohmian particles cannot be straightforwardly bunched together to form a macroscopic object, as can be done with Newtonian particles because the former exhibit a strong non-local behavior independent of the inter-particle separation. Hence, at some point, a quantum-to-classical transition has to be invoked to account for the disappearance of non-local effects at the macroscopic level. The problem is that the details of such a transition mechanism are far from being clear in a primitive ontology de Broglie-Bohm setting (see, e.g., \citealp{509}, for a preliminary proposal in this sense). Now, as already explicitly acknowledged in the analysis of emergence of subsystems in \S\ref{sec:dbb}, it is certainly true that also the transition from \eqref{debohm} to \eqref{new} needs further investigation to pinpoint the precise physical conditions that enable it, but this just means that the two approaches should currently be considered on a par regarding the quantum-to-classical transition issue.

\section{Conclusion: Open Questions and Future Developments}\label{sec:disc}

The discussion carried out in this paper shows how the treatment of non-relativistic quantum systems in PSD and its metaphysical reading are already fairly well-established. The de Broglie-Bohm framework is perhaps the easiest and conceptually cleanest environment for the development of a Leibnizian/Machian account of quantum phenomena that does not fall prey to the measurement problem. Likewise, the structural metaphysics that we propose accommodates and helps to clarify the relationship between entanglement structures and physical space---something that may pay huge dividends in the quantum gravity context. That being said, there are still a number of open issues in need of further inquiry. 

For starters, a number of technical and conceptual issues remain in our model. Firstly, as pointed out in \S\ref{sec:dbb}, our model does not satisfy the guidance equation at the universal level, which is definitely an important departure from standard de Broglie-Bohm theory. Crucially, as stressed in said section, our model does recover the predictions of standard QM for subsystems, including Born statistics, in the appropriate regime. However, the physical implications of this departure are still subject of investigation. Secondly, our current model demands a sound physical criterion for the classical limit. Two important issues are at play, in this case. First, the quantum coupling has been added to the dynamical system of our model by hand (recall the absence of Planck's constant in \eqref{debohm}). Although we provide a representation of this coupling by means of Bohr radii, we are currently looking into an explanation for its existence and the related emergence of Planck's constant. Second, we should supply a clear physical criterion for the regime in which the quantum potential becomes negligible.

These technical questions already represent a challenge to a full development of \eqref{debohm}. There is, however, another issue that concerns both the mathematics and the philosophy of quantum PSD models. This elephant in the room takes the form of $\Psi$ itself: As much as we can insist that $R(q)$ and $S(q)$ in \eqref{debohm} represent degrees of freedom inherent in the system itself---being just the way self-subsisting structures are held together by a world-building relation that possesses an entanglement component---the \emph{mathematical} structure representing $\Psi$ is external to any dynamical curve. Recalling how one of the major improvements of PSD over standard shape dynamics is the reduction of all dynamical aspects to geometric degrees of freedom of an unparametrized curve, it seems a bit odd to lightheartedly accept $\Psi$'s ``alien'' presence in shape space without feeling that PSD's reductionist spirit has been betrayed. The question then becomes: Is there any way to ``geometrize away'' $\Psi$? As we have seen in section \ref{sec:wf}, trying to cram $\Psi$'s whole mathematical structure into the geometry of  dynamical curves is a cumbersome task, to the point of technical intractability. But do we really need to reduce \emph{all} of the wave function's structure to geometric degrees of freedom? If we consider the ``gist'' of standard de Broglie-Bohm dynamics, we notice that the thing that drives the particles' motion is the gradient of the wave function's phase. This may suggest that we just need to incorporate this mathematical structure in the curves' geometry, which would then make the rest of $\Psi$ a collection of degrees of freedom that are redundant for the fundamental dynamics. This may open the door to showing that such extra structure---including Schr\"odinger's equation---appear in the framework as added layers of description from some sort of ephemeris construction that resembles the constructions used to recover notions such as scale and duration. This suggestion represents a possible solution to the problem, but it would require going beyond the model \eqref{debohm} since, as we have seen, the general dynamical scheme in this case involves much more than just the gradient of the wave function's phase to work. At the present stage, this route remains open for future investigations.

%it would require a process of mathematical reduction for $\Psi$ that goes well beyond a simple restriction of the non-relational wave function to shape space. 

Speaking of future lines of research, an important point to be stressed is that, although the de Broglie-Bohm framework is very friendly to PSD's reductionist procedure to construct relational models, this is by no means the only choice to construct quantum models. There is in principle no obstacle to construct, for example, GRW-like models of PSD---both in its flash and matter-density versions \citep{211,208}. Notoriously, GRW models carry counter-intuitive physical consequences, such as material objects being ``galaxies'' of sparse flashes, or even possessing space-like separated pieces resulting from the ``tails'' of the collapsed wave function. It would be very interesting to explore the metaphysical picture induced by these phenomena in a relational setting.

\pdfbookmark[1]{Acknowledgements}{acknowledgements}
\begin{center}
\textbf{Acknowledgements}:
\end{center}
AV and PN gratefully acknowledge financial support from the Polish National Science Centre, grant No. 2019/33/B/HS1/01772.

\bibliography{biblio}

\end{document}